\documentclass{aa}

\usepackage[varg]{txfonts}
\usepackage{xcolor}
\usepackage[normalem]{ulem}
\usepackage{comment}

\usepackage[symbol]{footmisc}

\usepackage{hyperref}
\hypersetup{
    colorlinks,
    linkcolor={blue!50!black},
    citecolor={blue!50!black},
    urlcolor={blue!50!black}
    }

\usepackage[encapsulated]{CJK}
\usepackage{ucs}
\newcommand{\orcid}[1]{\protect\href{https://orcid.org/#1}{\protect\includegraphics[width=8pt]{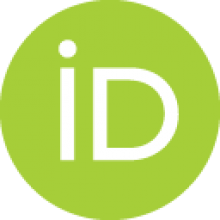}}}
\usepackage[utf8]{inputenc} 

\newcommand{\orcit}[1]{\protect\href{https://orcid.org/#1}{\protect\includegraphics[width=8pt]{orcid-ID.png}}}

\usepackage{comment}

\begin{document}

\makeatletter
\renewcommand{\linenumbers}{\relax}

\title{Helical magnetic field structure in 3C\,273 }
\subtitle{A Faraday rotation analysis using multi-frequency polarimetric VLBA data}

\author{
Teresa Toscano\orcid{0000-0003-3658-7862}\inst{\ref{IAA}}$^{,\dagger}$
\and Sol N. Molina\orcid{0000-0002-4112-2157}\inst{\ref{IAA}}
\and José L. Gómez\orcid{0000-0003-4190-7613}\inst{\ref{IAA}}
\and Ai-Ling Zeng\orcid{0009-0000-9427-4608}\inst{\ref{IAA}}
\and Rohan Dahale\orcid{0000-0001-6982-9034}\inst{\ref{IAA}}
\and Ilje Cho\orcid{0000-0001-6083-7521}\inst{\ref{KASI},\ref{Yonsei},\ref{IAA}}
\and Kotaro Moriyama\orcid{0000-0003-1364-3761}\inst{\ref{IAA}}
\and Maciek Wielgus\orcid{0000-0002-8635-4242}\inst{\ref{IAA}}
\and Antonio Fuentes\orcid{0000-0002-8773-4933}\inst{\ref{IAA}}
\and Marianna Foschi\orcid{0000-0001-8147-4993}\inst{\ref{IAA}}
\and Efthalia Traianou\orcid{0000-0002-1209-6500}\inst{\ref{IAA}}
\and Jan Röder\orcid{0000-0002-2426-927X}\inst{\ref{IAA}}
\and Ioannis Myserlis\orcid{0000-0003-3025-9497}\inst{\ref{IRAM}, \ref{Mpifr}}
\and Emmanouil Angelakis\orcid{0000-0001-7327-5441}\inst{\ref{Germany}}
\and Anton Zensus\orcid{0000-0001-7470-3321}\inst{\ref{Mpifr}}
}

\institute{
Instituto de Astrof\'{i}sica de Andaluc\'{i}a-CSIC, Glorieta de la Astronom\'{i}a s/n, E-18008 Granada, Spain\label{IAA} $^{\dagger}$\email{ttoscano@iaa.es}
\and Korea Astronomy and Space Science Institute, Daedeok-daero 776, Yuseong-gu, Daejeon 34055, Republic of Korea\label{KASI} 
\and Department of Astronomy, Yonsei University, Yonsei-ro 50, Seodaemun-gu, Seoul 03722, Republic of Korea\label{Yonsei} 
\and Institut de Radioastronomie Millimétrique, Avenida Divina Pastora, 7m Loca 20, E-18012 Granada, Spain\label{IRAM}
\and Max-Planck-Institut für Radioastronomie, Auf dem Hügel 69, D-53121 Bonn, Germany\label{Mpifr}
\and Orchideenweg 8, D-53123 Bonn, Germany\label{Germany}
}

\titlerunning{Helical magnetic field structure in 3C\,273}
\authorrunning{T. Toscano et al.}

\date{Received \today / Accepted \today}

\abstract{We present a study on rotation measure (RM) of the quasar 3C\,273. This analysis aims to discern the magnetic field structure and its temporal evolution. The quasar 3C\,273 is one of the most studied active galactic nuclei due to its high brightness, strong polarization, and proximity, which enables resolving the transverse structure of its jet in detail. We used polarized data from 2014, collected at six frequencies (5, 8, 15, 22, 43, 86 GHz) with the Very Long Baseline Array, to produce total and linear polarization intensity images, as well as RM maps. Our analysis reveals a well-defined transverse RM gradient across the jet, indicating a helical, ordered magnetic field that threads the jet and likely contributes to its collimation. Furthermore, we identified temporal variations in the RM magnitude when compared with prior observations. These temporal variations show that the environment around the jet is dynamic, with changes in the density and magnetic field strength of the sheath that are possibly caused by interactions with the surrounding medium.}


\maketitle

\section{Introduction}
\label{sec:intro}

Magnetic fields are considered a dominant factor in the launching, acceleration, and collimation of relativistic jets in Active Galactic Nuclei (AGN) \citep{Blandford_2019}. Theoretical models suggest that jet formation is driven either by the extraction of energy from a rotating black hole (the Blandford-Znajek mechanism; \citealt{Blandford_1977}) or by magnetic forces acting on material in the accretion disk (the Blandford-Payne mechanism; \citealt{Blandford_1982}). Both processes require the presence of a structured magnetic field with poloidal (along the jet axis) and toroidal (wrapped around the jet) components. Simulations using general relativistic magnetohydrodynamics (GRMHD)  and RMHD \citep[e.g.,][]{Fuentes_2018} have shown that the accretion flow around the black hole naturally develops a ``spine-sheath'' configuration, which may act as a magnetized boundary layer for the jet \citep{Asada2008, Gomez_2008, gomez_2012}.

Despite the success of numerical simulations in supporting these models, observational evidence for the detailed structure of the magnetic field, particularly its toroidal component, remains scarce \citep{Zamaninasab_2013, Molina_2014, Pasetto_2021}. Observations of linearly polarized synchrotron radiation provide insight into the magnetic field orientation projected on the plane of the sky. The highest resolution linear polarization maps are obtained from Very Long Baseline Interferometry (VLBI) observations \citep[e.g.,][]{Boccardi_2017, Hada_2019, Wielgus_2022, Jorstad_2023}, which probe the innermost regions of AGN jets. However, on their own, these polarization maps cannot fully constrain on their own the three-dimensional (3D) structure of the magnetic field.

To infer the magnetic field component along the line of sight (perpendicular to the plane of the sky), we take Faraday rotation into account. This effect consists on a rotation (dependent on the wavelength) of the polarization angle of the emission going through a magnetized medium. That is to say, the electric vector position angle (EVPA) is rotated by the (line-of-sight) magnetic field. Thus, the strength and direction of this magnetic field can be inferred through the Rotation Measure (RM), quantifying this rotation and providing information on the magnetized environment of the jet.

Theoretically, the RM is directly proportional to the line-of-sight magnetic field strength ($B_{||}$) and the electron density ($n_e$) in the medium, integrated along the path length (d$y$) \citep{Burn_1966}:

\begin{equation}
    RM = 8.1 \times 10^{5} \int{n_{e} B_{||} \text{d}y}
\end{equation}

where RM is measured in radians per square meter ($\text{rad} / \text{m}^{2}$), $B_{||}$ is in Gauss, $n_{e}$ is in $\text{cm}^{-3}$, and d$y$ is in parsecs (pc). Thus, by mapping the RM across a jet, we can investigate the magnetic field structure using different features such as gradients. For example, a smooth transverse gradient in a jet, given that RM is sensitive to the magnetic field's direction, would indicate a continuous change in the magnetic field's direction, which would evidence the existence of a helical magnetic field \citep{Blandford_1988, Blandford_1993, Asada_2002, gomez_2012, Gomez_2016, Pasetto_2021}.

In the simplest case, assuming a uniform magnetic field external to the jet \citep{Hovatta_2012}, the observed EVPA $\chi_{\rm obs}$ is related to the intrinsic EVPA $\chi_0$ and the square of the observing wavelength ($\lambda^2$) through the RM as:

\begin{equation}
    \chi_{\rm obs} = \chi_0 + RM \lambda^{2}
\end{equation}

The quasar 3C\,273, at redshift $z = 0.158$ \citep{Schmidt_1963}, is an ideal candidate for RM studies due to its brightness and high polarization, also showing in many cases a well-resolved jet structure \citep[e.g.,][]{Davis_1985, Conway_1993, Jester_2005, Perley_2017}. Over the years, it has been the subject of numerous VLBI monitoring campaigns aimed at understanding its jet morphology and magnetic field structure \citep[e.g.,][]{Krichbaum_1990, Lobanov_2001, Savolainen_2006, Kovalev_2016, Bruni_2017, Lister_2019, Lister_2021}.

The first detection of a transverse RM gradient in 3C\,273 was reported by \citet{Asada_2002}, suggesting the presence of a toroidal magnetic field. Later studies have confirmed this finding \citep{Zavala_2005, Asada_2005, Attridge_2005}, with similar RM gradients observed as well in other AGNs \citep[e.g.,][]{gomez_2012, Gabuzda_2017, Kravchenko_2017}, further supporting the notion that jets are threaded by helical magnetic fields \citep[e.g.,][]{Gomez_2016}.

In this paper, we present new multi-frequency VLBI observations of 3C\,273, focusing on the study of RM maps to better understand the geometry of the magnetic field structure in this source. We compare our results with previous works, offering new insights into the jet's magnetic field structure and evolution. The paper is organized as follows. Section \ref{sec:Obs} describes the observations, data reduction, and calibration procedures. Section \ref{sec:Data_analy} presents our results, which are then discussed in detail in Section \ref{sec:dicussion}. We also include an appendix with the goodness of the fit of the RM maps. Throughout this work, we assume a cosmology with $H_0 = 71$ km s$^{-1}$ Mpc$^{-1}$, $\Omega_m = 0.27$, and $\Omega_\Lambda = 0.73$ \citep{Komatsu_2009}, where 1 milliarcsecond corresponds to 2.71 pc in projected distance.

\section{Observation and data analysis}
\label{sec:Obs}

The observations presented in this work were conducted on 2014, November 21, using the Very Long Baseline Array (VLBA) as part of an astrometric multi-frequency program (with reference BG216). This program targeted a sample of BL~Lacs, flat spectrum radio quasars, and radio galaxies. Due to the astrometric nature of the program, the observing strategy involved alternating scans between the target sources and nearby calibrators at different frequencies. As a result, the scans on the target source were relatively short -typically less than 30 seconds - an approach standard in astrometric experiments to maintain high positional accuracy while minimizing systematic errors introduced by atmospheric or instrumental effects. This strategy, combined with interleaving scans at multiple frequencies, provided comprehensive $uv$-coverage for the sources.

All 10 VLBA antennas were present during the observation, which spanned six frequencies: 4.98, 8.42, 15.25, 21.89, 43.79 and 87.55 GHz.
The data were recorded in 8 IFs of 32 MHz, amounting to 256 MHz total bandwidth per frequency. 
The on-source time ranged from 10 minutes at the lower three frequencies, to 4h 46m at 22GHz, and 47m at the highest two frequencies.

\subsection{Data reduction}

The calibration and processing of the data were performed using the National Radio Astronomy Observatory (NRAO) Astronomical Image Processing System (AIPS; \citealt{Greisen_2003}). Standard procedures for polarimetric VLBI observations were followed for phase and amplitude calibration, as described in the AIPS Cookbook\footnote{\href{http://www.aips.nrao.edu/CookHTML/CookBook.html}{http://www.aips.nrao.edu/CookHTML/CookBook.html}}. We applied a priori calibration to the correlated visibility amplitudes using system temperature measurements and gain curves specific to each station. Phase calibration was carried out by performing a fringe-fitting procedure on the data, after correcting for the parallactic angle. Additionally, we corrected for phase bandpass, delay, and rate offsets, which resulted in strong fringe detections across all baselines.

Subsequent data editing was performed interactively using Difmap software \citep{Shepherd_1994, Shepherd_2011} to remove outliers and poor quality data, which could introduce noise or artifacts into the imaging process. The final images were produced through iterative CLEAN deconvolution, ensuring that the imaging process accurately reconstructed the brightness distribution of the source while minimizing noise.

\subsection{Polarization and Rotation Measure Calculation} \label{subsec:pol_rot}

The instrumental polarization calibration (commonly referred to as D-term calibration) was performed using the task LPCAL in AIPS, following the correction for cross-hand delays using the RLDLY task. Once the data are fully calibrated both in intensity and polarization, to calculate the rotation measure it is necessary to first calibrate the absolute electric vector position angle at each frequency by using single-dish measurements or close-in-time observations from monitoring programs. For 15 and 43 GHz, we obtained the EVPA correction using archival data from the MOJAVE \citep{Lister_2009} and BEAM-ME \citep{Jorstad_2017} monitoring programs. We applied the appropriate rotation by comparing our observed EVPAs with these archival values.

For 5 and 8 GHz, we used single-dish data from the QUIVER (Monitoring the Stokes Q, U, I and V Emission of AGN jets in Radio) project, obtained with the Effelsberg radio telescope \citep{Myserlis_2018, Angelakis_2019} in the frame of the F-GAMMA monitoring program. These observations, conducted at 4.85 and 8.35 GHz, were taken 10 days after our VLBA session and yielded EVPA values of $-28 \pm 2^\circ$ and $-36 \pm 2^\circ$, respectively. For 22 GHz, no single-dish measurements were available, hence to find the appropriate rotation used for the RM map we first corrected 15 and 43 GHz polarization images, we calculated the average EVPA of each map and then, assuming the linear relation mentioned in \autoref{sec:intro}, we interpolated the average for the 22 GHz map in order to apply the corresponding rotation. In this way, we provide a reasonable approximation in the absence of direct data.

The RM maps were generated using data from three frequencies at a time, ensuring consistent $uv$-coverage, pixel size, and image resolution across the different frequencies. For an appropriate comparison between frequencies, higher-frequency images were convolved with the restoring beam of the lowest frequency. Only regions with polarized intensity with a signal-to-noise ratio greater than 4$\sigma$ (where sigma is the root mean squared (rms) value from Difmap) were used in the analysis. If no polarization was detected simultaneously at all three frequencies, the corresponding RM pixel was blanked in the map.


Image alignment across different frequencies was necessary due to the core shift caused by synchrotron self-absorption \citep{Blandford_1979}. We aligned the images using two-dimensional cross-correlation, following the method described by \citet{Walker_2000} and \citet{Croke_2008}. After cross-correlation, the cumulative shifts for each frequency relative to the reference (lowest frequency) were determined and corrected for. 
Finally, due to the inherent $\pi$ ambiguity in the EVPAs, $\pi$ rotations were manually applied to ensure the best fit to the $\lambda^2$ law in the RM analysis.

The VLBA antennas are not optimized for observations at 86 GHz, resulting in poor sensitivity at this frequency, both in total intensity and polarized flux. Because of this limited sensitivity, the lack of significant polarization as well as the absence of single-dish measurements available for calibrating the absolute EVPA, the 86 GHz image was not included in the Faraday rotation analysis.

\begin{figure*}
\centering 
\includegraphics[width=0.83\columnwidth]{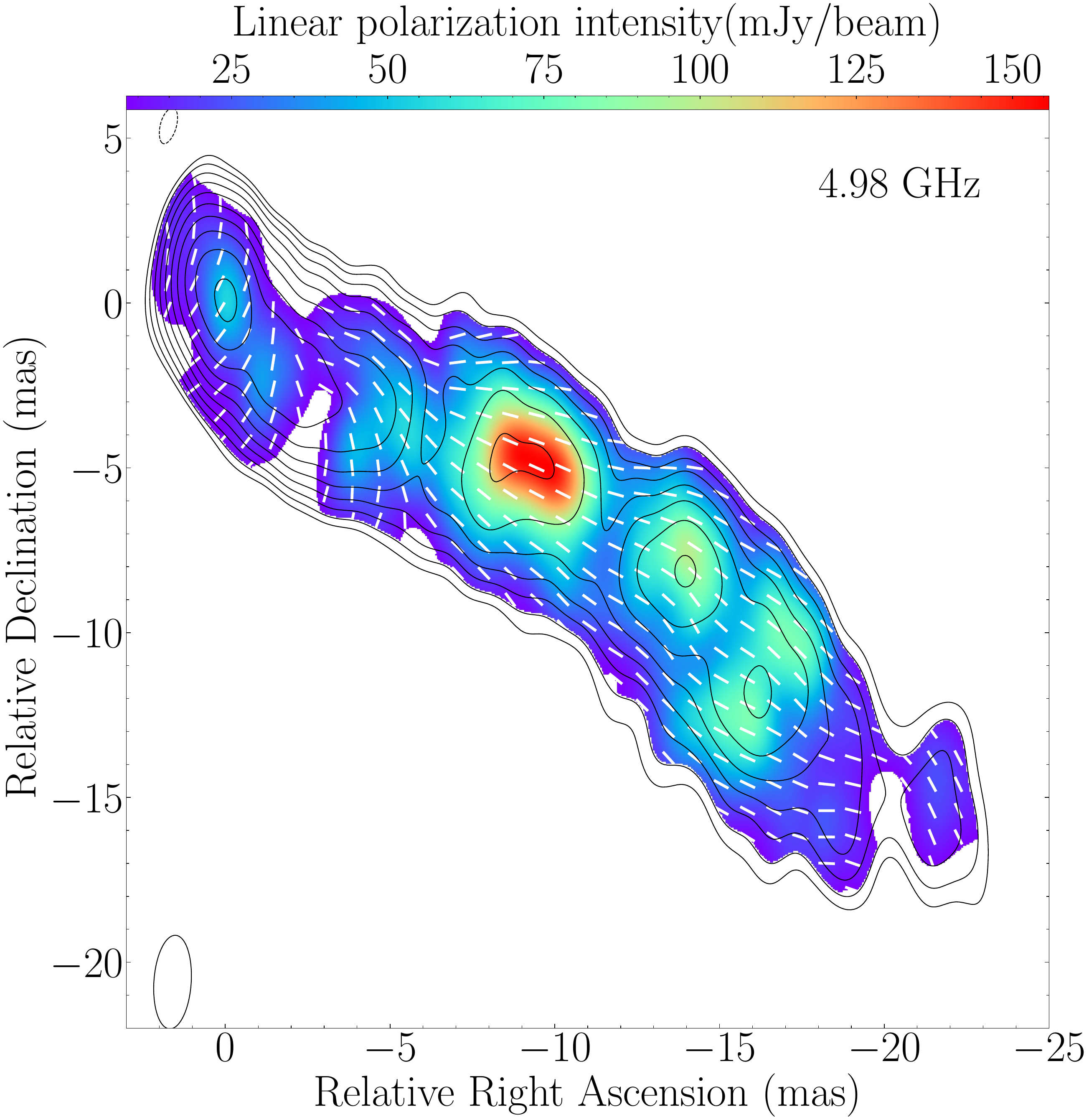}
\includegraphics[width=0.83\columnwidth]{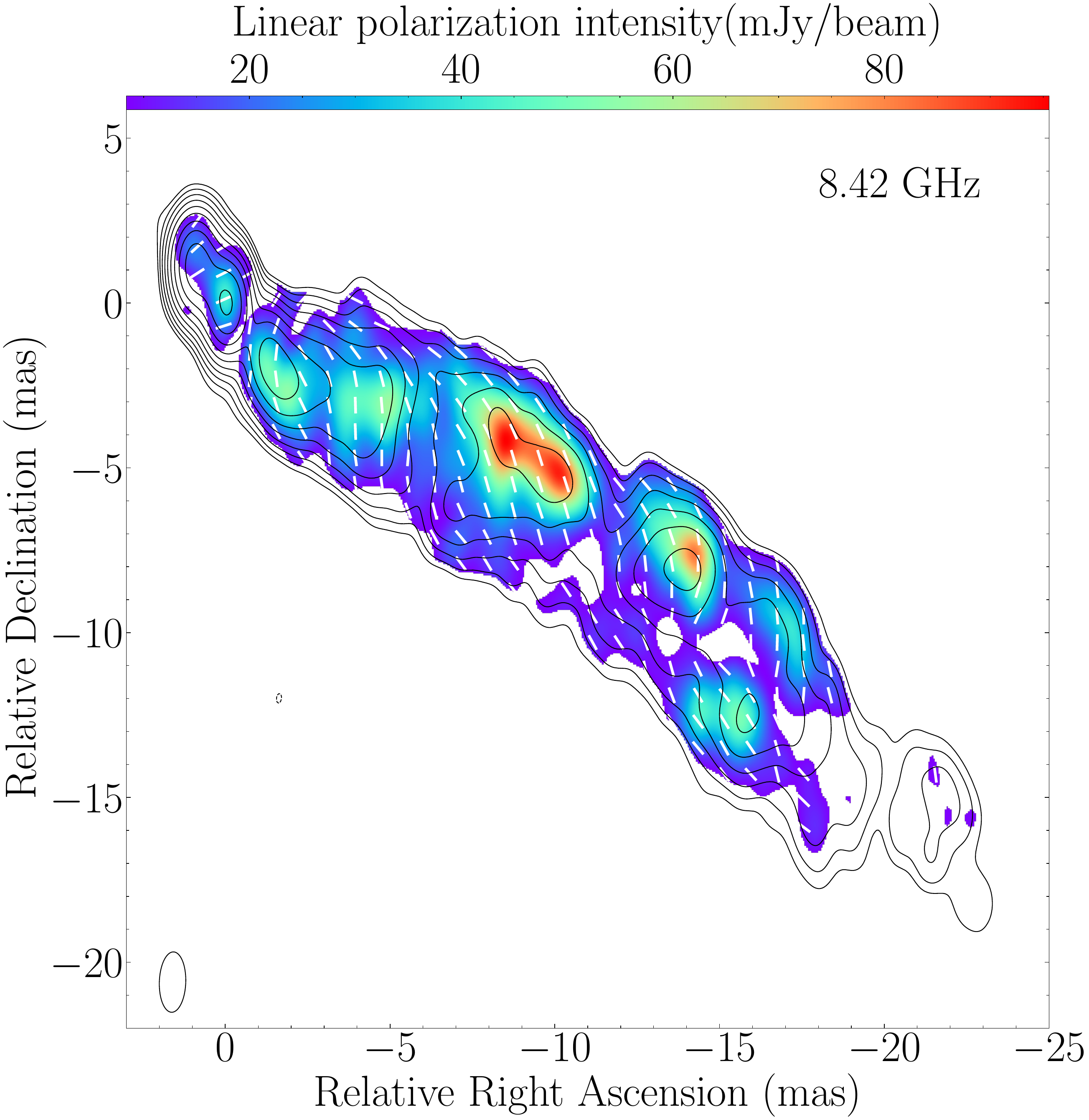} \\
\includegraphics[width=0.83\columnwidth]{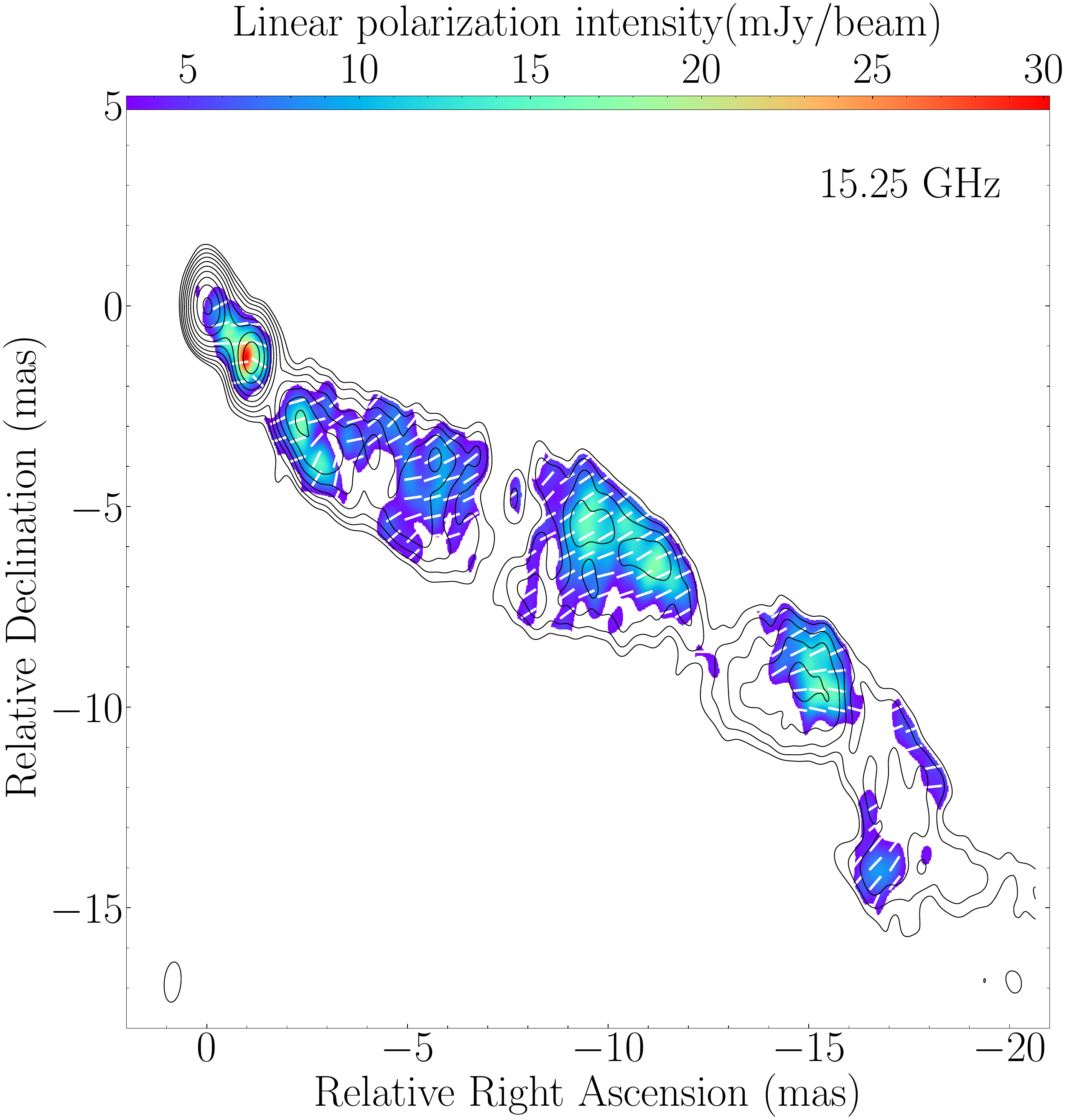}
\includegraphics[width=0.83\columnwidth]{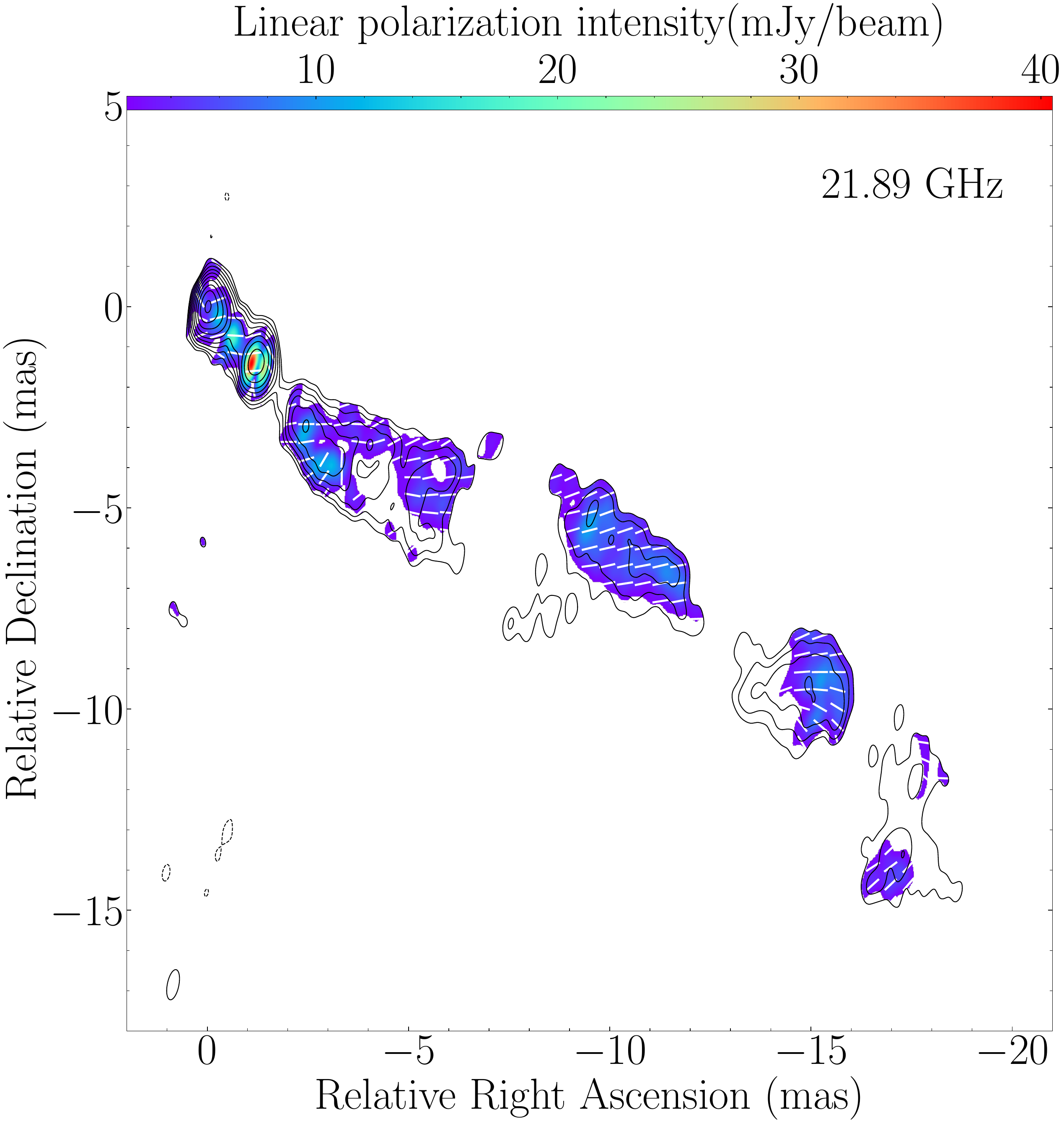} \\
\hspace{0.3cm}\includegraphics[width=0.83\columnwidth]{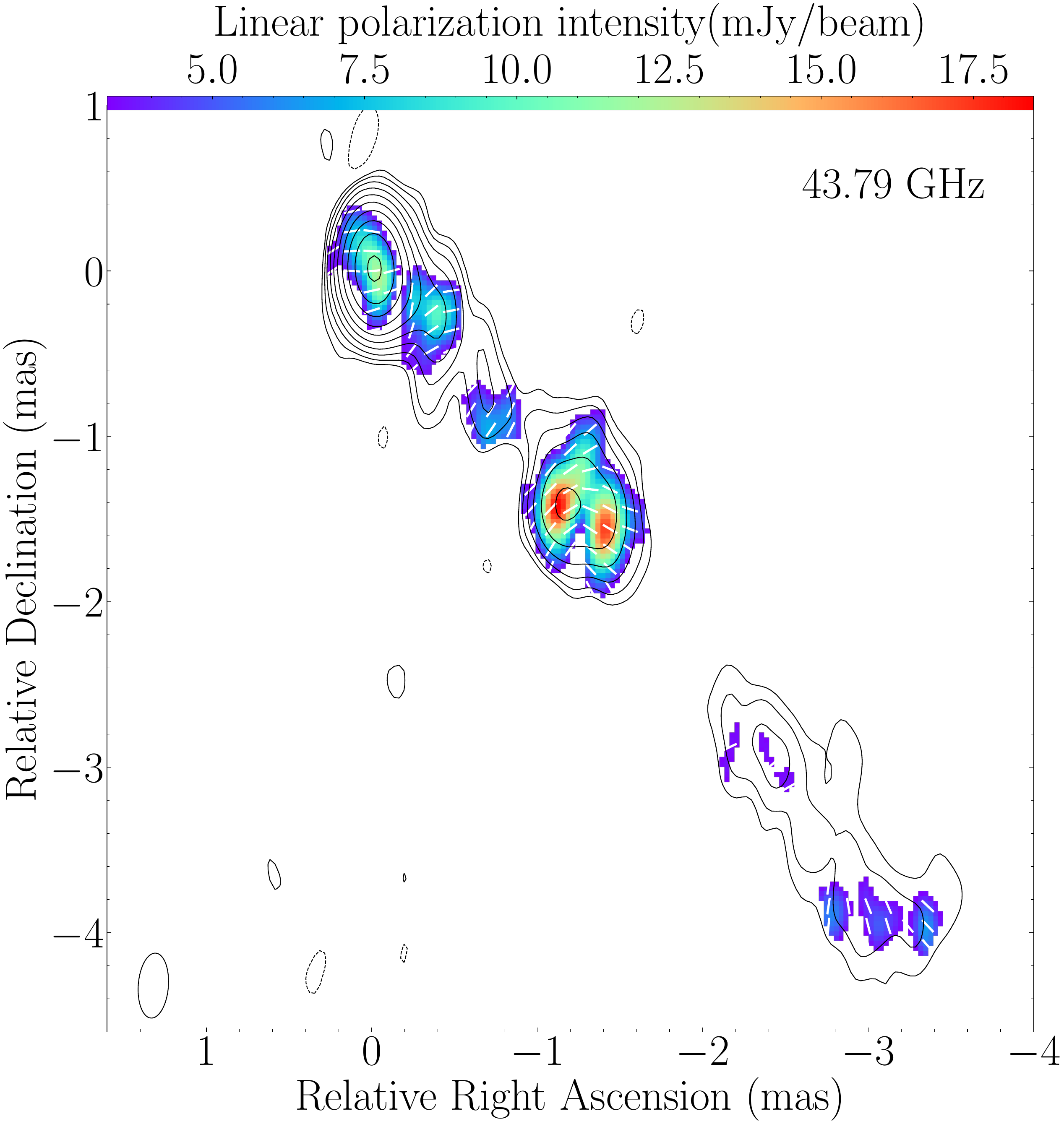}
\includegraphics[width=0.85\columnwidth]{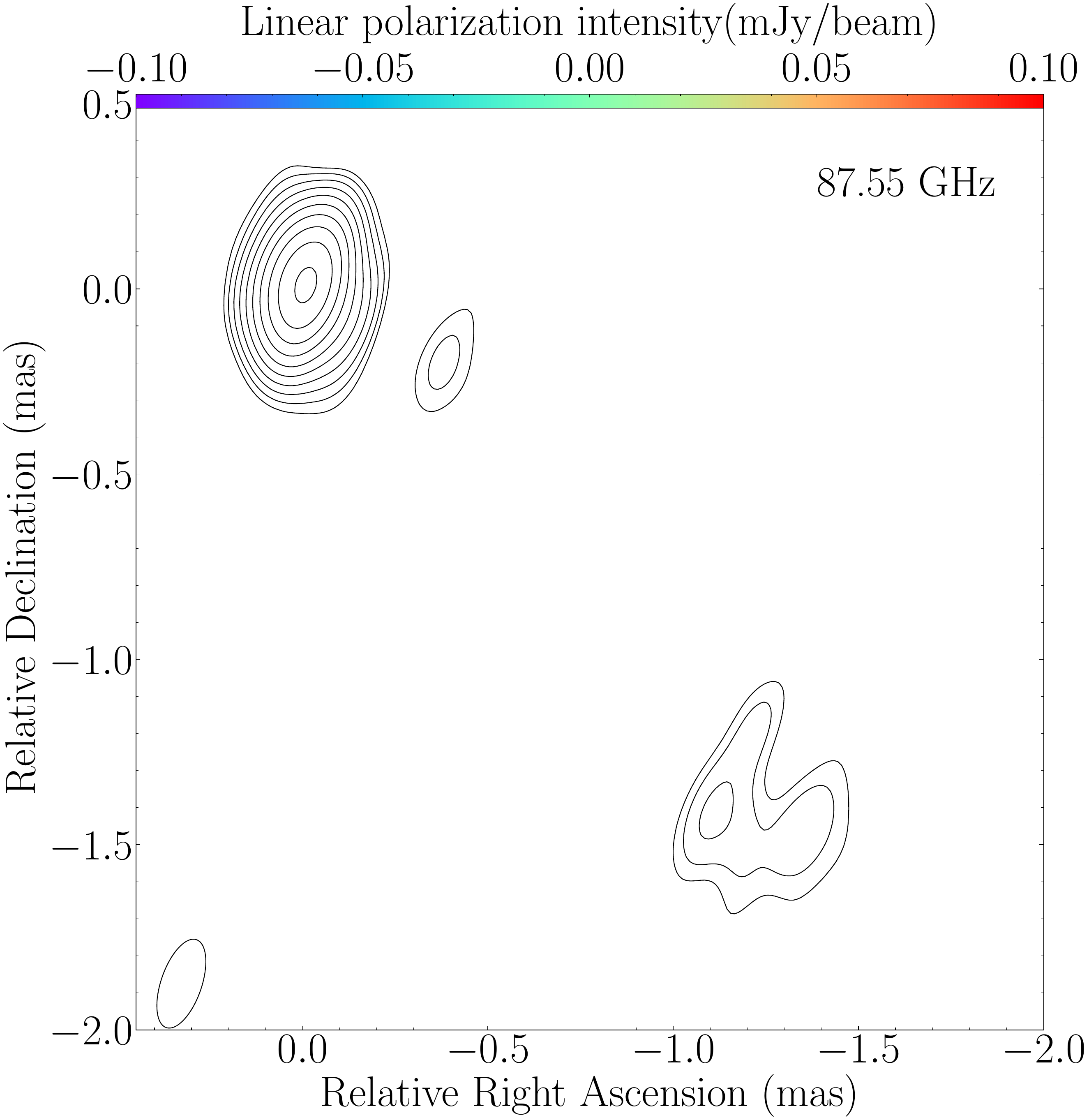} \\
\caption{VLBA total and linearly polarization images at different frequencies. Contours show total intensity above a threshold of 5$\sigma$ level, color scale represents the linearly polarized intensity above 4$\sigma$ level, and white ticks indicate the observed electric vector position angle.}
\label{fig:linear_pol}
\end{figure*}



\section{Results}
\label{sec:Data_analy}

In this section, we present VLBA polarimetric images of 3C\,273 at 5, 8, 15, 22, 43, and 86\,GHz (\autoref{fig:linear_pol}). The total intensity images are shown with contour lines above the $5\sigma$ level, while the EVPAs are overlaid as white ticks. The total polarized intensity is displayed in color, with only pixels above $4\sigma$ included in the polarization map; any pixels below this threshold are blanked.

We provide two RM maps derived from observations at 5-8-15\,GHz and 15-22-43\,GHz, which allow us to investigate the magnetic field structure across different regions of the jet (\autoref{fig:rm_map}). Finally, we present images showing transverse sections across the jet at several locations to analyze the asymmetry of the jet in different aspects: the total intensity, the linear polarization intensity and the degree of polarization at 5 GHz (\autoref{fig:I_ridgeline_cuts}), and the RM distribution in both RM maps (\autoref{fig:I_ridgeline_cuts_2}).

\subsection{Maps of Linear Polarization}
\label{subsec:Linear_pol_1}
\begin{figure*}[ht!]
\centering 
\includegraphics[width=2.1\columnwidth]{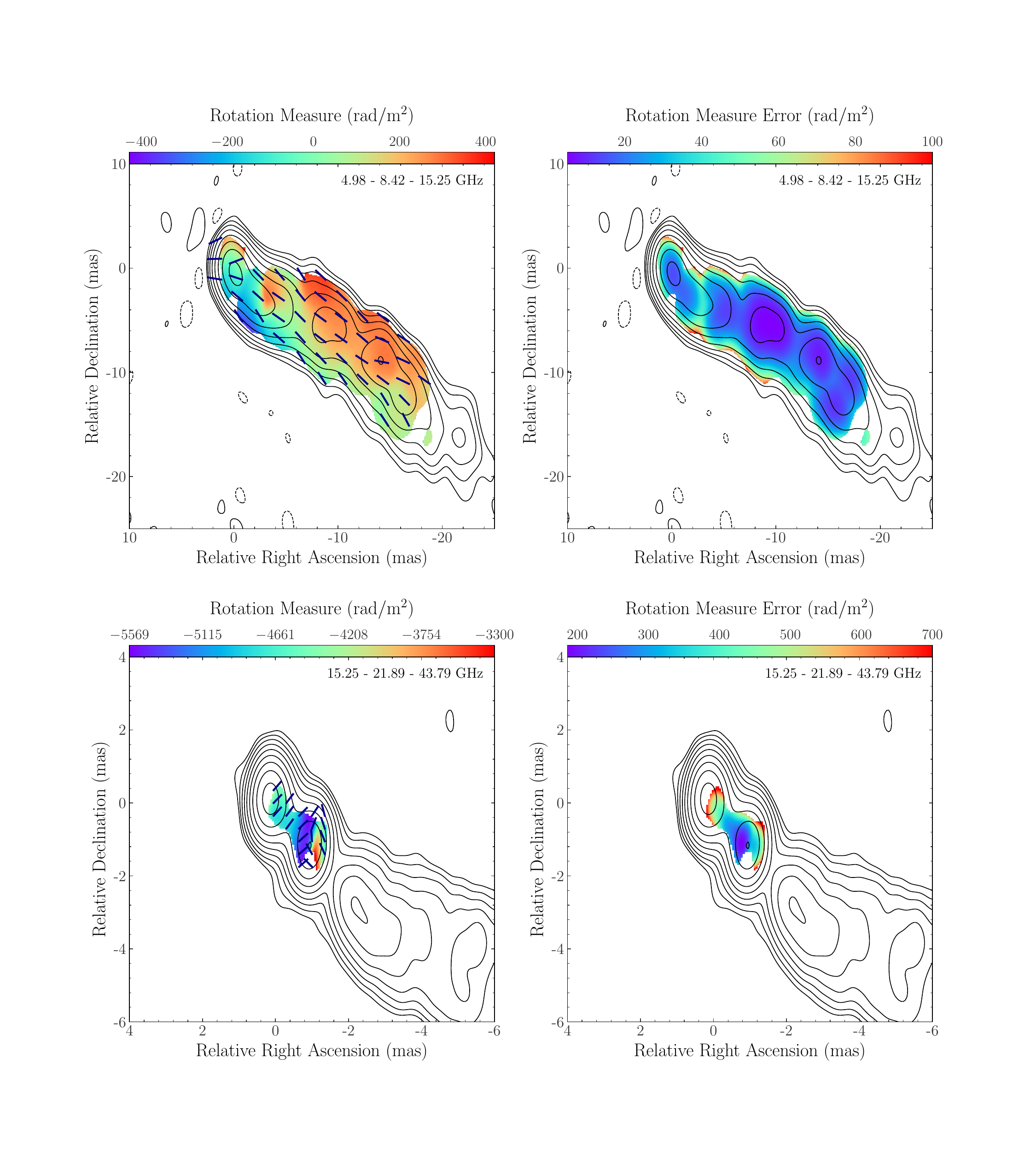}

\caption{Rotation measure maps of 3C\,273 using 5-8-15\,GHz (left column, top) and 15-22-43\,GHz (left column, bottom) and its respective errors (right column). Corrected (intrinsic) EVPAs are plotted as dark blue ticks.
}
\label{fig:rm_map}
\end{figure*}

The total intensity and polarization structure in \autoref{fig:linear_pol} of 3C\,273 at different frequencies are consistent with previous observations of 3C\,273 \citep{Asada_2002, Zavala_2005, Hovatta_2012}.
As discussed in \autoref{subsec:RM_map_1} (see also \autoref{fig:rm_map}), the EVPAs are significantly influenced by Faraday rotation, and once corrected, the magnetic field appears to be predominantly toroidal.

The core is located at the northeastern, upstream end of the jet, where linear polarization is significantly reduced, primarily due to opacity and beam depolarization effects. At 5 and 8 GHz, a bright, highly polarized component is visible around 10 mas from the core. This feature fades at higher frequencies, although the overall jet structure remains consistent with archival images from the MOJAVE and BEAM-ME programs. As the core becomes progressively optically thin at 15 and 22 GHz, a new component emerges around 1 mas from the core. Close-in-time MOJAVE observations at 15 GHz and BEAM-ME observations at 43 GHz indicate that this component is moving outward. A comparison with \citet{Hada_2016}, which used 43 GHz data from February 2014, shows the presence of this component at 1 mas, consisting of two subcomponents referred to as P2 and P3. In our polarization images taken nine months later, these P2 and P3 components appear to evolve and move further downstream from the core.

We also present one of the few images of 3C\,273 at 86 GHz, which is consistent with previous observations by \citet{Attridge_2005}, \citet{Hada_2016}, \citet{Hovatta_2019}, and more recently by \citet{Hiroki_2022}, which incorporates ALMA data for the first time.  Although the sensitivity at this high frequency is limited, the total intensity image plotted in contours reveals the core and the brightness structure (component)around 1 mas, consistent with lower-frequency images. Unfortunately, no significant polarization is detected at this frequency.

\subsection{Rotation Measure Maps}
\label{subsec:RM_map_1}

In this section, we present Rotation Measure maps in \autoref{fig:rm_map} derived from two frequency sets: 5-8-15 GHz and 15-22-43 GHz, along with their corresponding error maps. Previous RM studies \citep{Zavalatailor_2001, Wardle_2018, Hovatta_2019} typically utilized only two frequencies for a linear fit. By incorporating three frequencies in our analysis, we achieve a more robust and statistically significant estimate of RM. The goodness of the fit of the linear relation is presented in Appendix \ref{sec:appendix}, where we have used the reduced $\chi^{2}$ as a metric, as used in \cite{Lisakov_2021}, blanking pixels that had $\chi^{2} > 5.99$ (with 3 data points and 2 degrees of freedom indicates a 95\% confidence level).
The maps, shown in \autoref{fig:rm_map}, are overlaid with total intensity contours and the RM-corrected EVPAs. The RM error maps have been constructed using the square root of the variance of the parameter estimates from the linear fitting.

For the first RM map, derived from 5-8-15 GHz observations, the values range from approximately -400 rad m$^{-2}$ to 400 rad m$^{-2}$, which are consistent in order of magnitude with respect to those reported in previous studies at similar frequencies \citep{Asada_2002, Asada2008, Hovatta_2012, Lisakov_2021}. A clear transverse RM gradient is evident, presenting also a sign change across the jet. This gradient reflects a systematic variation in the magnetized plasma's properties across the jet, consistent with the geometry and structure expected for a persistent helical magnetic field.

As a byproduct of obtaining the RM maps from a linear approximation, we estimate the intrinsic EVPA of the source (i.e. after correcting for Faraday rotation), which corresponds to the y-intercept of the linear regression. Plotted as dark blue ticks, they follow the jet direction, signaling that the Faraday rotation effect is significant (see \autoref{fig:linear_pol} for comparison) and indicating a predominant toroidal component of the magnetic field structure.

In contrast, the second RM map, based on 15-22-43 GHz observations, is limited to the innermost 2~mas of the jet. At these higher frequencies, the core region is closer to the base of the jet, where the line of sight passes through a denser, more strongly magnetized plasma. Consequently, the RM values increase significantly (around one order of magnitude), with absolute values exceeding $5 \times 10^{3}$ rad m$^{-2}$. Such high RM values are common at higher frequencies, with values exceeding $2 \times 10^{4}$ rad m$^{-2}$ reported at 3 and 7 mm \citep{Attridge_2005, Hada_2016}, and even reaching $5 \times 10^{5}$ rad m$^{-2}$ in some cases \citep{Hovatta_2019}, suggesting a dense Faraday screen or stronger magnetic fields \citep{Savolainen_2008}. The larger negative values of the RM seen in the higher frequencies RM are consistent with the lower frequencies RM map, showing also negative RM values for this region. The clear RM gradient seen in the lower frequencies RM is not present in the 15-22-43 GHz map. Furthermore, as observed also in the lower frequencies RM map, the intrinsic EVPAs in the 15-22-43 GHz RM map appear almost perpendicular to the jet direction, becoming more aligned with the jet at around 2 mas. Note that this is based on the assumption already discussed in Sec. \ref{subsec:pol_rot}.

\begin{figure*}[ht!]
\centering 
\includegraphics[width=1.01\columnwidth]{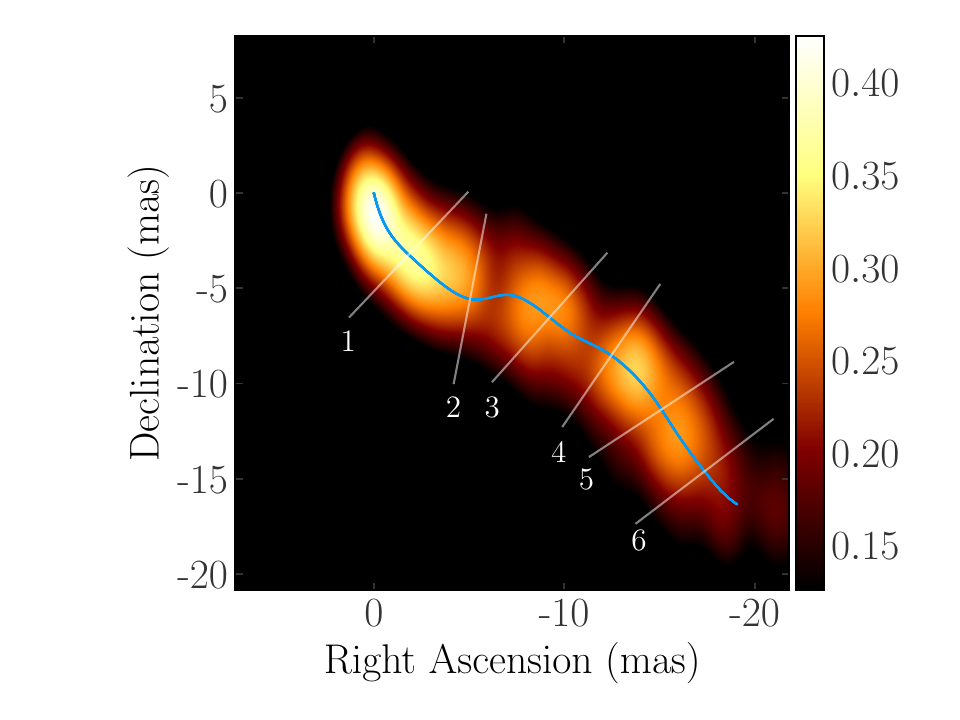}
\includegraphics[width=0.98\columnwidth]{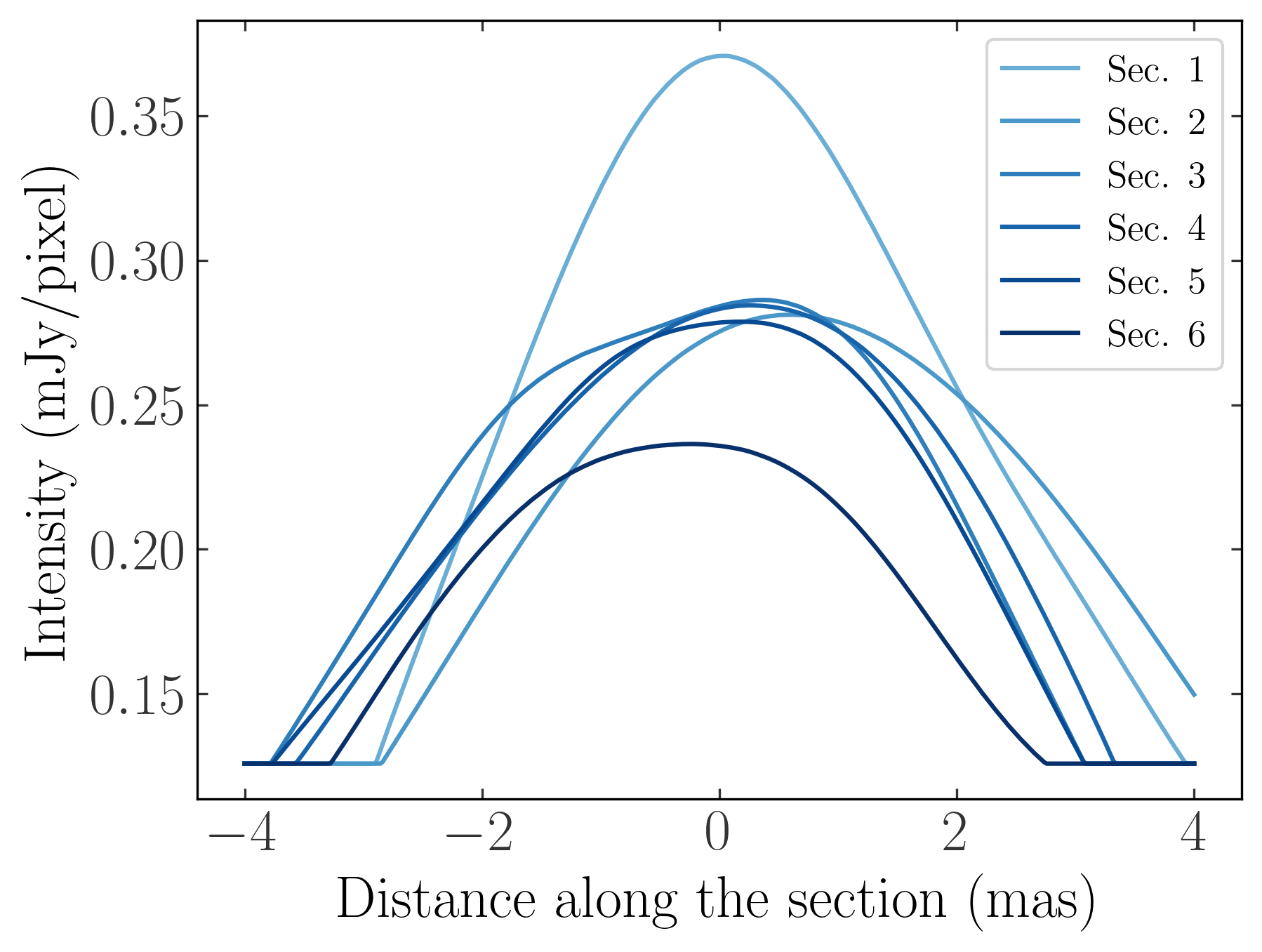}

\includegraphics[width=\columnwidth]{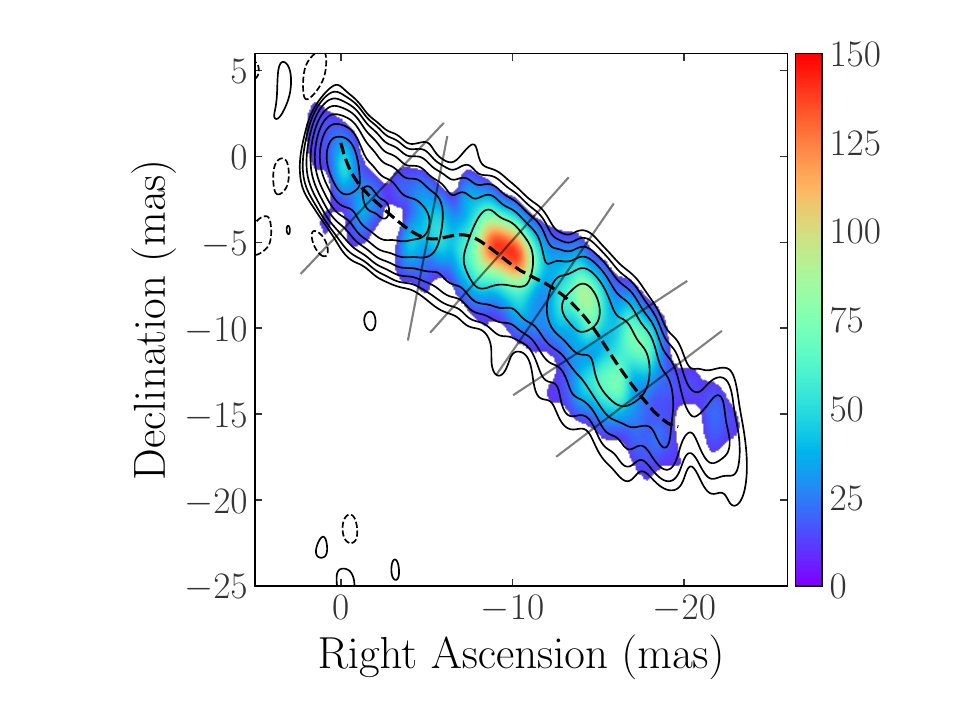}
\includegraphics[width=\columnwidth]{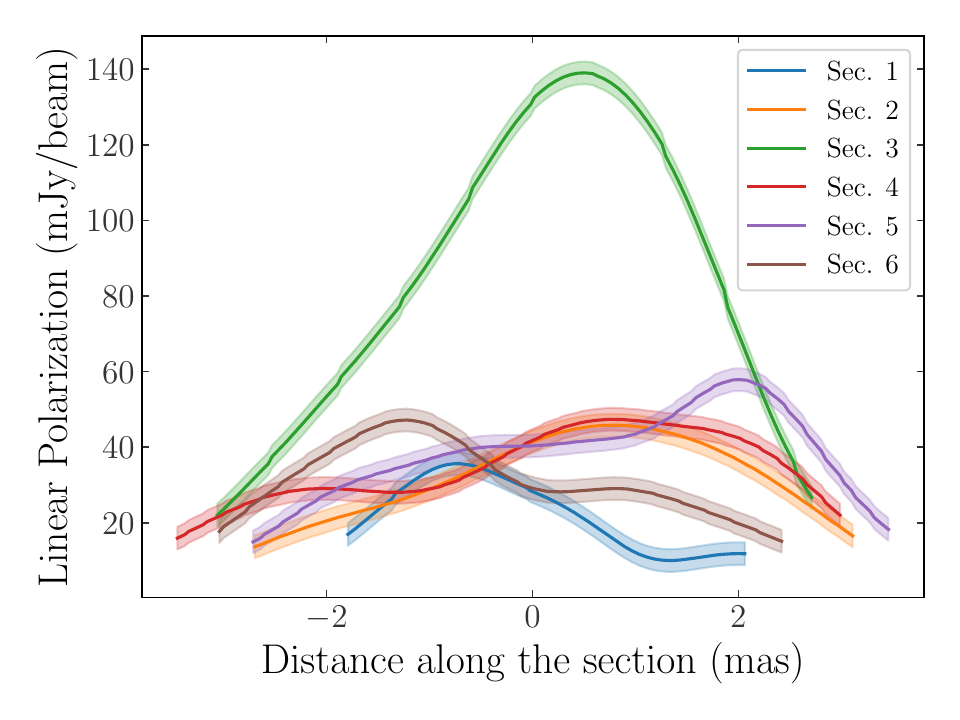}       

\includegraphics[width=\columnwidth]{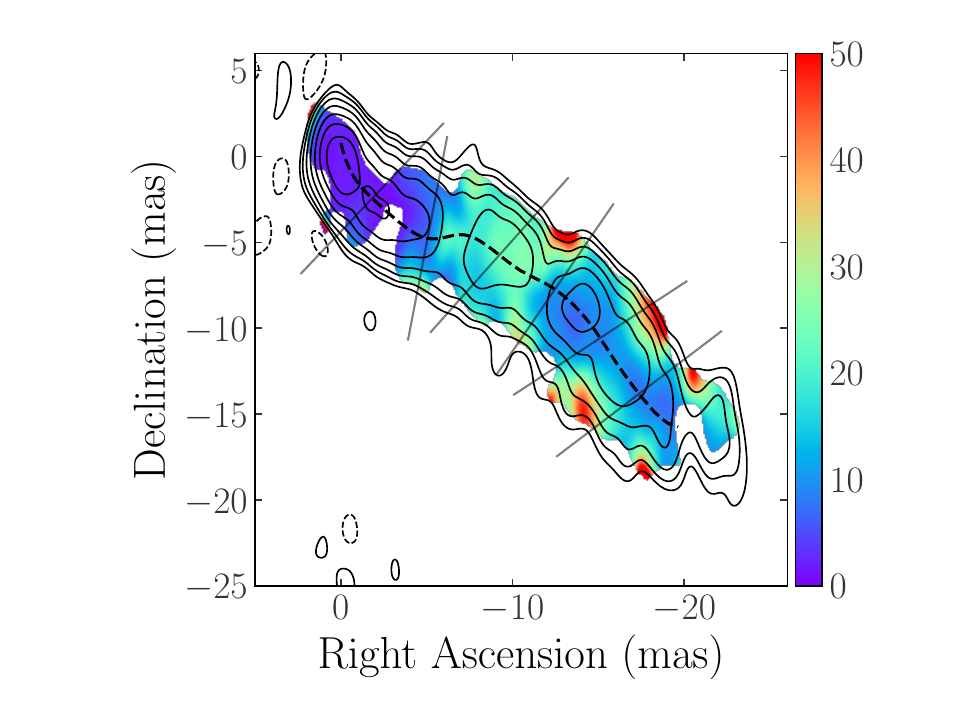}
\includegraphics[width=\columnwidth]{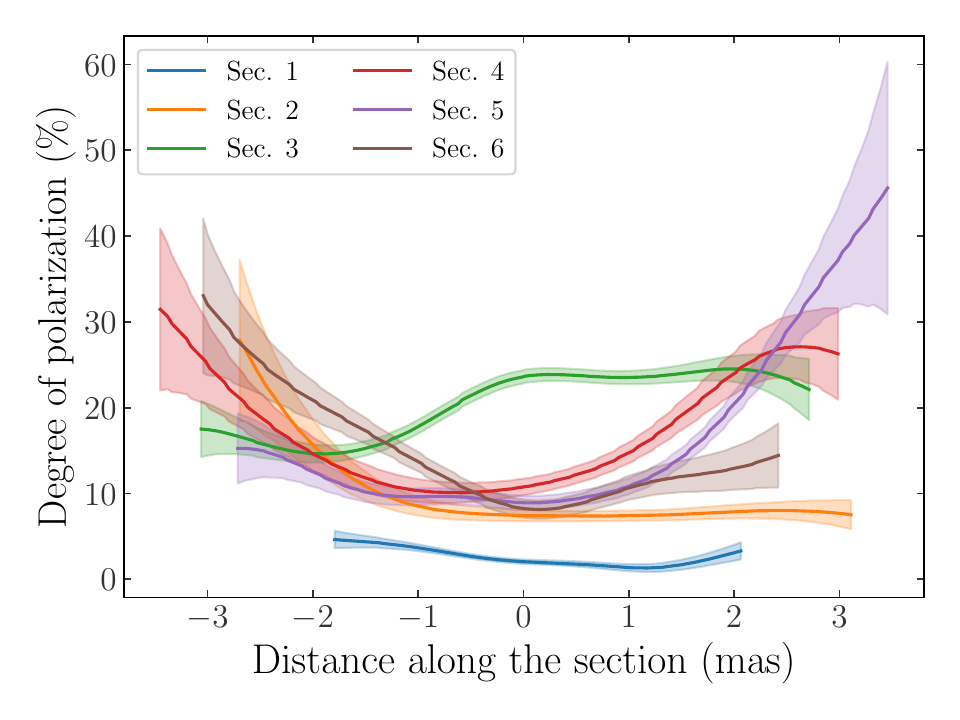}

\caption{Transversal sections for total intensity image at 5 GHz (top row), linear polarization at 5 GHz (central row) and degree of polarization (bottom row). Units displayed in the left column. Order of the sections along the jet are numbered in the total intensity image.}

\label{fig:I_ridgeline_cuts}
\end{figure*}

\begin{figure*}
\centering 

\includegraphics[width=\columnwidth]{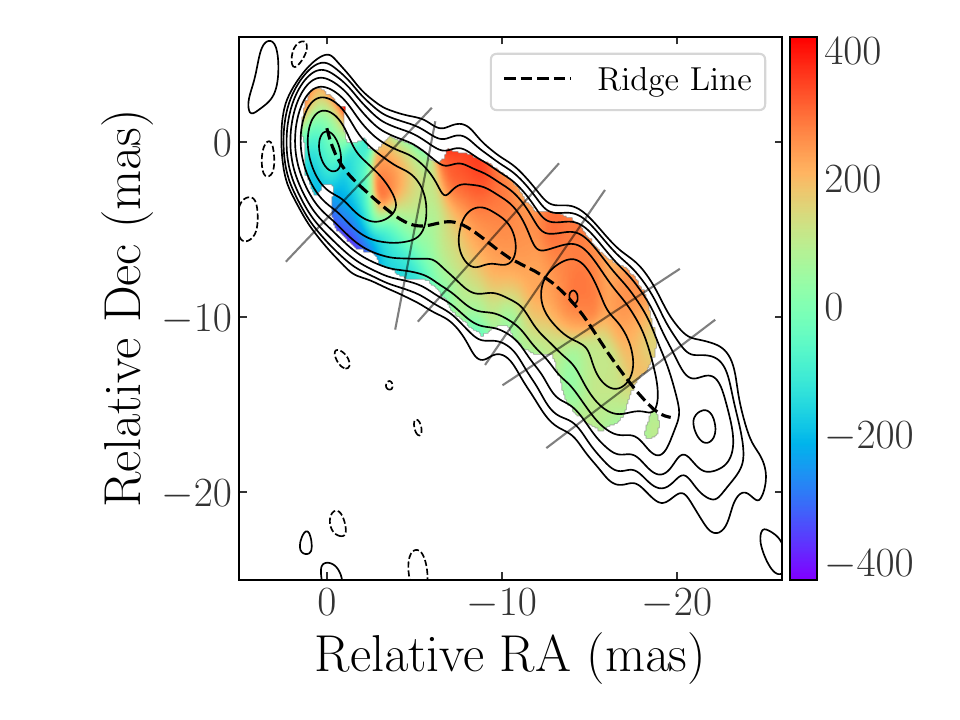}
\includegraphics[width=1.02\columnwidth]{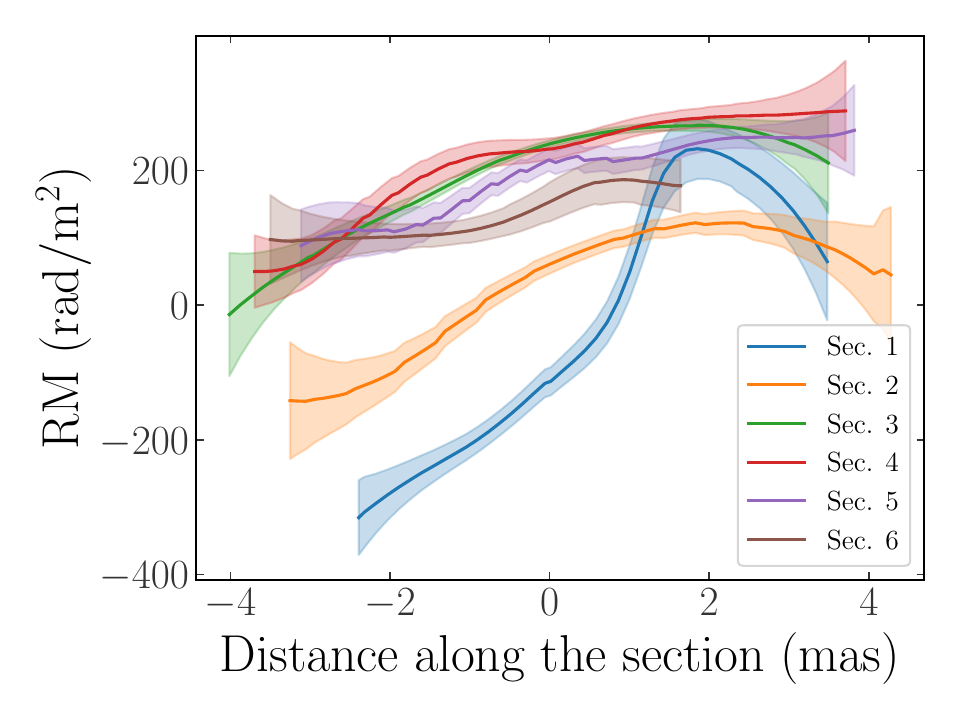}

\includegraphics[width=\columnwidth]{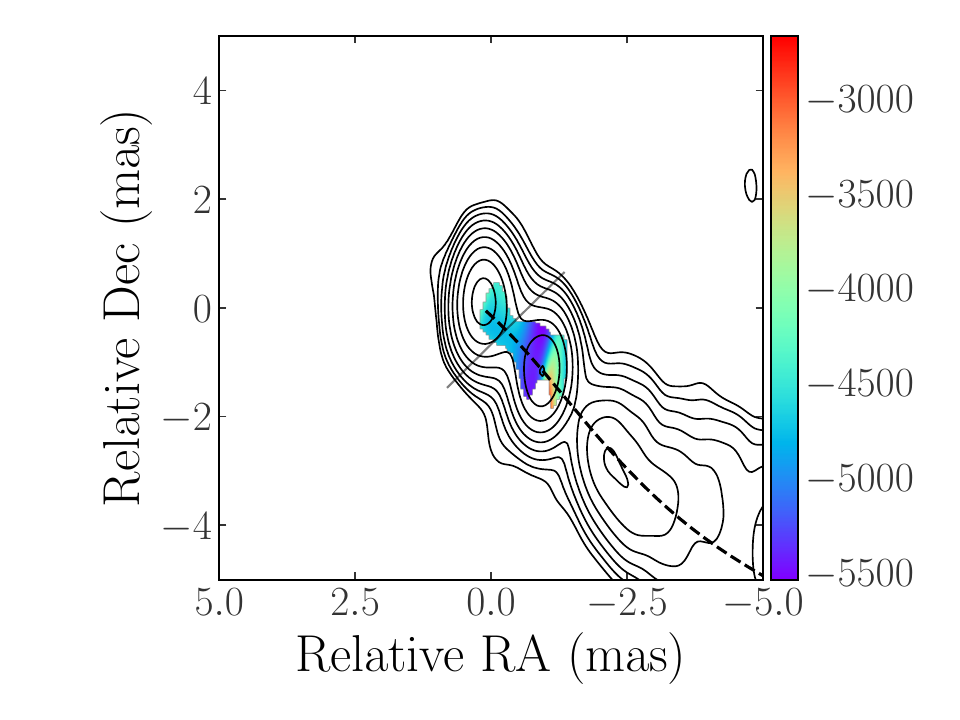}
\includegraphics[width=1.02\columnwidth]{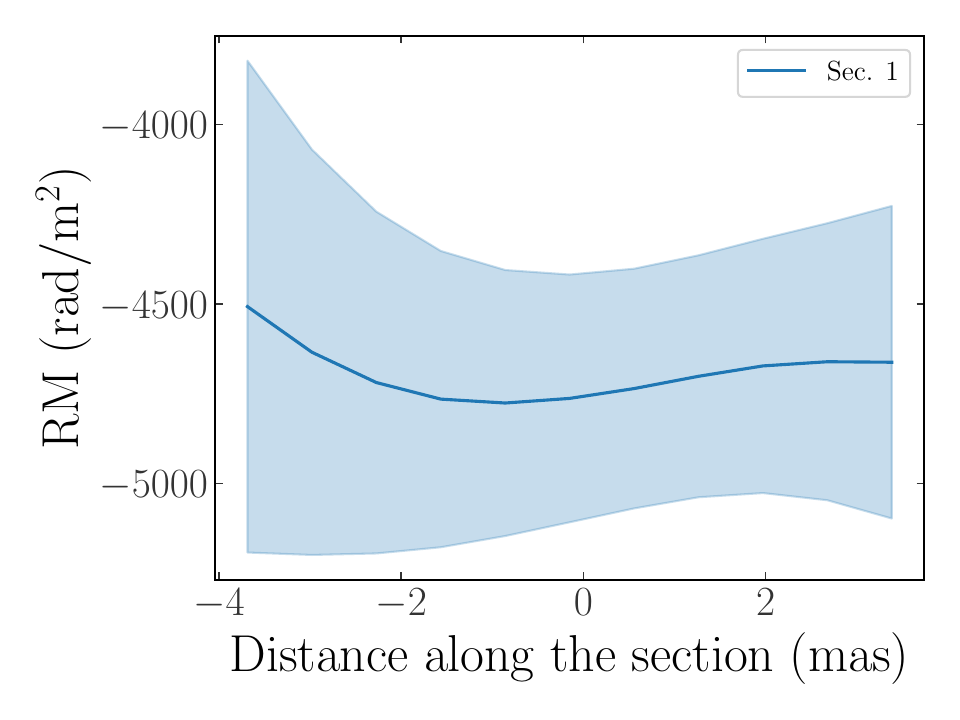}

\caption{Transversal sections for RM map using 5-8-15 GHz (top row) and 15-22-43 GHz (bottom row). Order and values of the sections are displayed from left to right, as shown in \autoref{fig:I_ridgeline_cuts}}

\label{fig:I_ridgeline_cuts_2}
\end{figure*}

\subsection{Asymmetry Across the Jet and the existence of a helical magnetic field}

To analyze the transverse jet asymmetry, we trace a ridge line (curve that traces the total intensity along the jet axis) following the method of \citet{Fuentes_2023}. We apply six evenly spaced transverse sections along this ridge line to examine variations in total intensity, linear polarization and degree of polarization along the length of the jet. \autoref{fig:I_ridgeline_cuts} shows the jet sections (units of the colorbars are displayed in the y-axis of the right column plots). 
In addition, we also display a second study for rotation measure maps in \autoref{fig:I_ridgeline_cuts_2}, using the same six sections mentioned above for the 5-8-15 GHz RM map, and only one section in the 15-22-43 GHz RM map, coincident with the one closest to the core in the lower frequency RM map.

In \autoref{fig:I_ridgeline_cuts}, the intensity image in the first row reveals a clear drop in emission from the first section near the core to the subsequent sections downstream, reflecting the expected decrease in brightness as we move away from the core region. In most of the sections, the northern part of the jet appears consistently brighter than the southern part, suggesting a slight asymmetry in the intensity distribution. This asymmetry, aligns with predictions from relativistic magnetohydrodynamic simulations of jets threaded by a helical magnetic field \citep{Fuentes_2018, Fuentes_2021}. 

The linear polarization map at 5 GHz displayed in the second row panels displays that overall, the jet shows more linearly polarized emission close to the spine, decreasing towards the borders. This behavior is especially relevant in the third section, where a highly polarized component is found at 10 mas away from the core, creating a spike reaching almost 140 mJy/beam.

In the degree of polarization map, we observe a stratification across the jet width, with systematically higher values towards the jet edges, as expected for the case in which the jet is threaded by a helical magnetic field \citep[e.g.,][]{Gomez_2008}. Only the third section in the degree of polarization shows higher values near the jet axis, most likely due to the presence of a strong shock leading to enhanced polarization, as seen in \autoref{fig:linear_pol}.

Moving on to the asymmetry of RM maps in \autoref{fig:I_ridgeline_cuts_2}, the first map (5-8-15 GHz, top row) shows a consistent transverse gradient in all sections, extending up to around 18 mas or a projected distance of 45 pc. The gradient is steeper near the core, where the stronger magnetic fields result in higher RM values. As we move further downstream, the asymmetry in RM persists, with higher positive values on the northern side and negative values on the southern side. The gradient weakens gradually as we move down the jet but remains detectable. These results agree with previous studies that report stronger RM magnitudes in the nuclear regions of active jets \citep{Asada_2002, Hovatta_2012}. The persistent RM gradient suggests a systematic change in the magnetic field along the line of sight.
For the second RM map at higher frequencies, the core region shows that the asymmetry is quite less obvious transversely, although there seems to change alongside the jet. The lack of further detection in this region prevents us from making a more thorough analysis.



\section{Discussion}
\label{sec:dicussion}

Our observations reveal a clear transverse RM gradient across the jet of 3C\,273 (see \autoref{fig:rm_map}) using 5, 8, and 15\,GHz, confirming that this feature persists both in time and along the jet. To contextualize our results, we compare them with previous RM measurements of 3C\,273, highlighting both consistencies and differences to map the jet evolution.

\citet{Asada_2002, Asada2008} first reported an RM gradient using observations from 1995 and 2002 at 4.6–8.6\,GHz. They measured exclusively positive RM values ranging from approximately 130 to 480\,rad\,m$^{-2}$. \citet{Zavala_2005} observed higher RM values using data from 2000 at 12–22\,GHz, with RM values ranging from about –50 to 350\,rad\,m$^{-2}$, predominantly positive. \citet{Wardle_2018} compiled these RM measurements and added those by \citet{Chen_2005}, based on 1999–2000 observations, which were positive and within a range of 200 to 800\,rad\,m$^{-2}$.

A significant finding by \citet{Hovatta_2012} was the detection of a sign change across the transverse RM gradient in 3C\,273, using data from 2006 at four frequencies between 8 and 15\,GHz. They suggested that this change was due to a different part of the jet being illuminated compared to earlier observations, similar to what was observed in 3C\,120 \citep{Gomez_2011}. Unlike \citet{Zavala_2005}, who observed mostly constant, predominantly positive RM values, \citet{Hovatta_2012} detected rapid RM variations within a three-month time span. The maximum gradient they observed ranged from about –600 to 500\,rad\,m$^{-2}$ at approximately 3–7 mas from the core.
This sign change was corroborated in a subsequent study by \citep{Lisakov_2021}, using data from 2009 at frequencies of 8.1–15.4\,GHz. Their RM values, ranging from –500 to 400\,rad\,m$^{-2}$, were consistent with those of \citet{Hovatta_2012}.

A possible explanation for the shift from exclusively positive RM values in earlier observations to a gradient with both negative and positive values in more recent measurements could be the variability in the jet's magnetic field configuration and/or the magnetized medium producing the Faraday rotation. Changes in the illuminated regions of the jet due to evolving emission patterns may cause different cross-sections to be observed, which in turn would be dominated with varying magnetic field orientations and polarization, thus maybe leading lead to the detection of both positive and negative RM values across the jet width. 

Additionally, variations in the internal conditions of the jet, such as the electron density or magnetic field strength could modify the effect of Faraday rotation. Also, interactions with the surrounding medium or even changes in the viewing angle due to the jet's curvature may also affect the observed RM values, whether modifying the external Faraday screen or altering the line-of-sight magnetic field. On another hand, new and more refined observational techniques or increased sensitivity could further reveal finer RM structures 
Therefore, the observed shift likely results from a combination of intrinsic changes in the jet's properties, environmental interactions, and changes in observational capabilities.

With our 2014 observations at six different frequencies, we examined both the structure and evolution of the polarization of 3C\,273, contributing to a more comprehensive understanding of its RM variations. As described in \autoref{subsec:RM_map_1}, our RM values range from –400 to 400\,rad\,m$^{-2}$, consistently demonstrating that the transverse gradient and the sign change are still present.

Contrasting our results with those closest in time, from five years earlier \citep{Lisakov_2021}, we observe minor fluctuations in the RM values, although it is complicated to establish an accurate comparison given that in this previous study the RM maps used only two sets of frequencies (4.5-8.4 GHz and 8.1-15.4 GHz) in the analysis. 

Both negative and positive values seem to remain within the same area of the jet, not showing a significant variation. We observe that negative RM values have since then slightly decreased in number even if their position has not noticeably changed, staying in the southern side of the jet up to $\sim$ 10 mas. The positive values, on the other hand, have interestingly increased in number, occupying a broader region of the jet and a bigger fraction of the jet width. 

The slight increase in the number of positive RM values and their positioning suggest a few possibilities: first, an evolving asymmetry in the density of the magnetized plasma surrounding the jet. This might indicate that either the electron density or the magnetic field strength has increased on the upper side of the jet over time. Second, the change could signal a structural variation in the jet, such as precession, slight bends or changes in viewing angle, altering how the magnetic field is perceived. This last option may be less probable due to the observed stability of the source structure over the years. 

The origin of the RM gradient in 3C\,273 remains a topic of investigation. Several models have been proposed to explain this phenomenon, broadly categorized into external and internal Faraday rotation scenarios. \citet{Asada_2002} suggested that the RM gradient arises from a helical magnetic field external to the jet, serving as an external Faraday screen. \citet{Zavala_2005} also argued for an external origin, likely a sheath surrounding the jet. They stated that if the high RM values were due to a uniform magnetic field in the sheath, severe depolarization would be expected, which was not observed. Later studies proposed that RM variations are due to changes in the external, slow-moving sheath, explaining RM changes on a timescale of several years \citep{Asada2008}. On the other hand, other suggested scenarios include as well internal Faraday rotation \citet{Hovatta_2012} based on observed rapid RM changes.
Similarly to \citet{Asada2008}, \citet{Lisakov_2021} proposed that the Faraday-rotating medium is an extensive sheath enveloping the jet, possibly consisting of slow-moving plasma that accounts for the observed RM stability and gradual variations over time. Our results are fully consistent with this model, suggesting that the screen has not changed significantly, especially between 2009 and 2014, and therefore supporting the idea of a relatively stable external Faraday screen. In addition, \citet{Lisakov_2021} predicted that if the jet direction changes farther to the south, mostly negative values will be observed, and since we observe that the positive values have increased, it seems that this is not the case. According to our observations, the jet direction is going towards the north, and might go back to have only positive values were it to continue that trend.
 
The observed EVPAs (see \autoref{fig:linear_pol}) are changing from parallel to almost perpendicular to the jet general direction as the frequency changes, while after correcting for the Faraday rotation effect (see \autoref{fig:rm_map}), they show to be mostly parallel to the jet (at lower frequencies). Not only this manifests the important effect of the Faraday rotation, but also indicates that the intrinsic magnetic field is predominantly perpendicular to the jet direction. That is, it has a strong toroidal component.
The presence of a predominantly toroidal magnetic field may contribute to stabilizing the jet flow, as evidenced by the relatively constant structure of this jet over the years, and play a crucial role in collimating the jet and maintaining its stability over large distances from the central engine.
The innermost region, though, shows that close to the core the rotation measure values are quite high and negative, and also the EVPAs appear to be rotated almost 90 degrees. This change in direction could be explained by optical thickness, which prevents us from fully visualize the magnetic field structure in the core region.

Polarization images from our observations show a component at around 10\,mas that displays an increase in polarization degree with increasing wavelength, probably due to a shock occurring in this region.
Additionally, we observe a gradient in the degree of polarization across the jet, with polarization values increasing towards the jet edges (\autoref{fig:I_ridgeline_cuts}, third row). In such a configuration, the magnetic field lines wrap around the jet axis, causing the magnetic field vectors to have different orientations across the jet width. Towards the jet edges, the magnetic field becomes more ordered and aligned perpendicular to the line of sight, resulting in higher observed degree of polarization. In contrast, at the jet center, the magnetic field may be more tangled or aligned along the line of sight, leading to lower polarization degrees.
Nevertheless, in the third section of the jet we observe that instead of a decrease, there is an increase of polarization degree. This is in agreement with the presence of a shock component, which would order or align the magnetic field in that region, therefore resulting in higher values.

Recent advancements made possible by highly sensitive radio telescopes, such as the Atacama Large Millimeter/submillimeter Array (ALMA) and global collaborations like the Event Horizon Telescope (EHT), have yielded significant new results at higher frequencies. These advancements provide high-resolution images that illuminate the collimation processes of the jet and its transition from a parabolic to a conical shape \citep{Casadio_2017, Hiroki_2022}. Future work involving new multi-frequency polarization as well as rotation measure studies of 3C 273 will surely widen our understanding of the behavior and evolution of this source.

Considering all the evidence, our results support the existence of a predominantly helical magnetic field in 3C\,273, consistent with previous studies \citep{Asada_2002, Zavala_2005, Asada2008, Attridge_2005, Hovatta_2012, Hovatta_2019}. The persistent and mostly stable transverse RM gradient, EVPA direction, together with an exhaustive comparison with literature provide evidence that the Faraday rotation is most likely external in nature, probably involving an external sheath enveloping the jet. Our findings strengthen the notion that large-scale, ordered magnetic fields play a crucial role in the collimation and stability of AGN jets \citep[e.g.,][]{Hovatta_2012, gomez_2012}.

By conducting further observations with improved sensitivity and resolution, we will be able to better characterize the magnetic field structure and the roles of internal and external Faraday rotation in AGN jets, contributing to a more comprehensive picture of the jet's magnetic field configuration and evolution, as well as the mechanisms underlying jet formation and stability in active galactic nuclei.

\begin{acknowledgements}
Author contributions: T. Toscano performed the analysis of total intensity, polarization, and Faraday rotation, and wrote most of the manuscript. J. L. Gómez wrote the observing proposal, prepared the schedule, calibrated the data with Sol M. Molina, contributed to the manuscript, and supervised the analysis. A. Zeng contributed to the rotation measure analysis and manuscript revision. R. Dahale provided scripts for plotting polarization images and commented on the final manuscript. I. Cho, K. Moriyama, M. Wielgus, A. Fuentes, M. Foschi, T. Traianou and J. Röder contributed to the discussion of results and revised the manuscript. I. Myserlis, E. Angelakis, and J. A. Zensus provided EVPA single-dish measurements at 5 and 8 GHz.\\

The work at the IAA-CSIC is supported in part by the Spanish Ministerio de Econom\'{\i}a y Competitividad (grants AYA2016-80889-P, PID2019-108995GB-C21, PID2022-140888NB-C21), the Consejer\'{\i}a de Econom\'{\i}a, Conocimiento, Empresas y Universidad of the Junta de Andaluc\'{\i}a (grant P18-FR-1769), the Consejo Superior de Investigaciones Cient\'{\i}ficas (grant 2019AEP112), and the State Agency for Research of the Spanish MCIU through the ``Center of Excellence Severo Ochoa" grant CEX2021-001131-S funded by MCIN/AEI/ 10.13039/501100011033 awarded to the Instituto de Astrof\'{\i}sica de Andaluc\'{\i}a.
\end{acknowledgements}

\bibliography{references}{}
\bibliographystyle{aa}

\appendix

\onecolumn

\section{Goodness of the linear fit for RM maps}
\label{sec:appendix}

\begin{figure*}[ht!]
    \centering
    \includegraphics[width=0.72\linewidth]{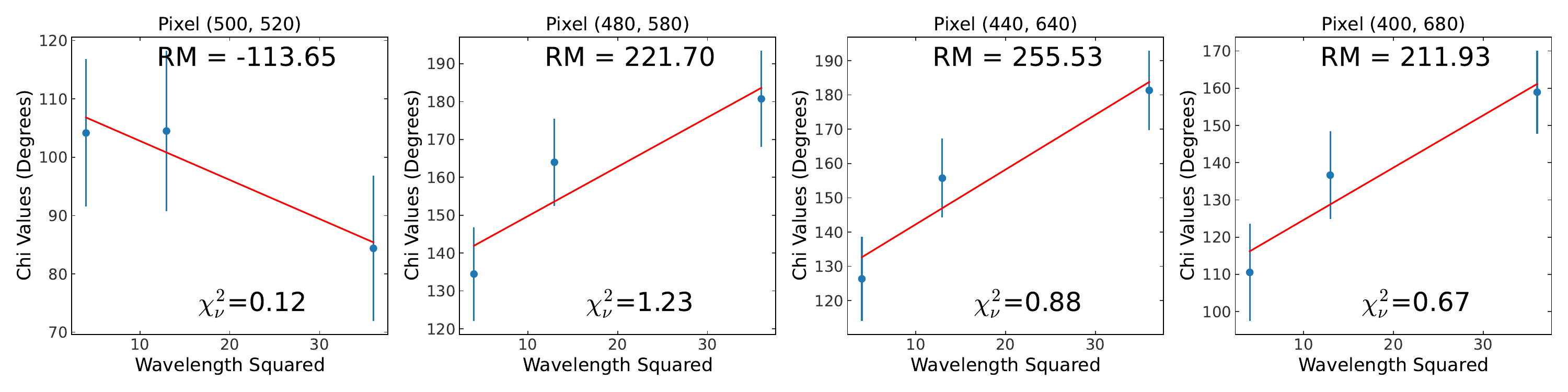}
    \includegraphics[clip, trim=10cm 0cm 03.5cm 0cm, width=0.21\linewidth]{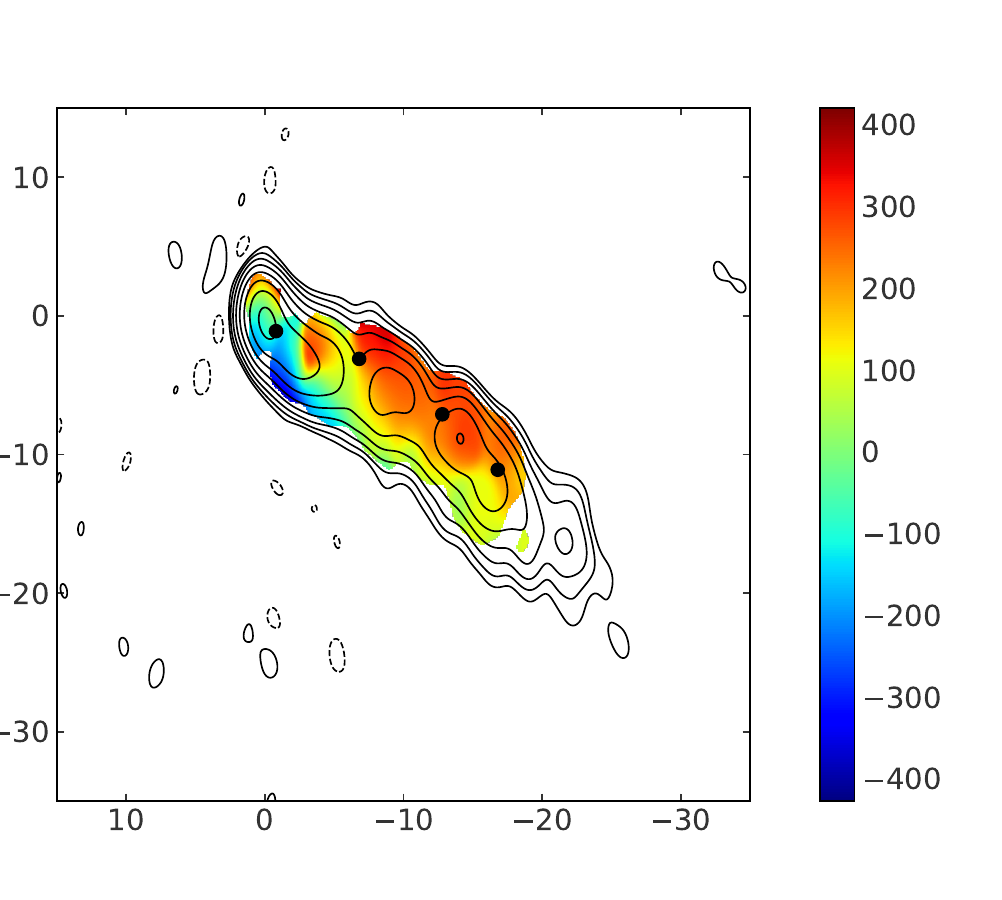}
    \includegraphics[width=0.72\linewidth]{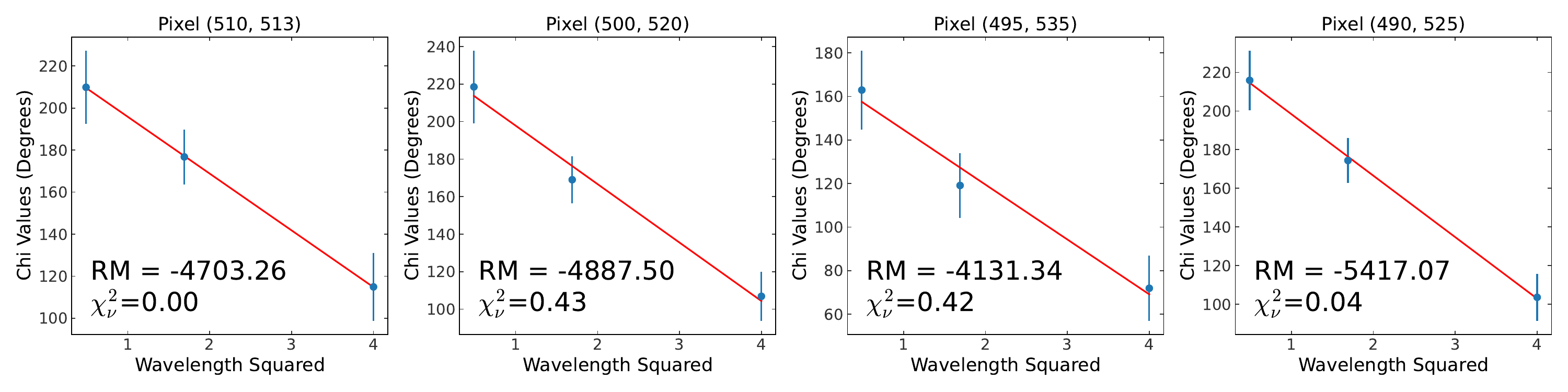}
    \includegraphics[clip, trim=10cm 0cm 03.5cm 0cm, width=0.21\linewidth]{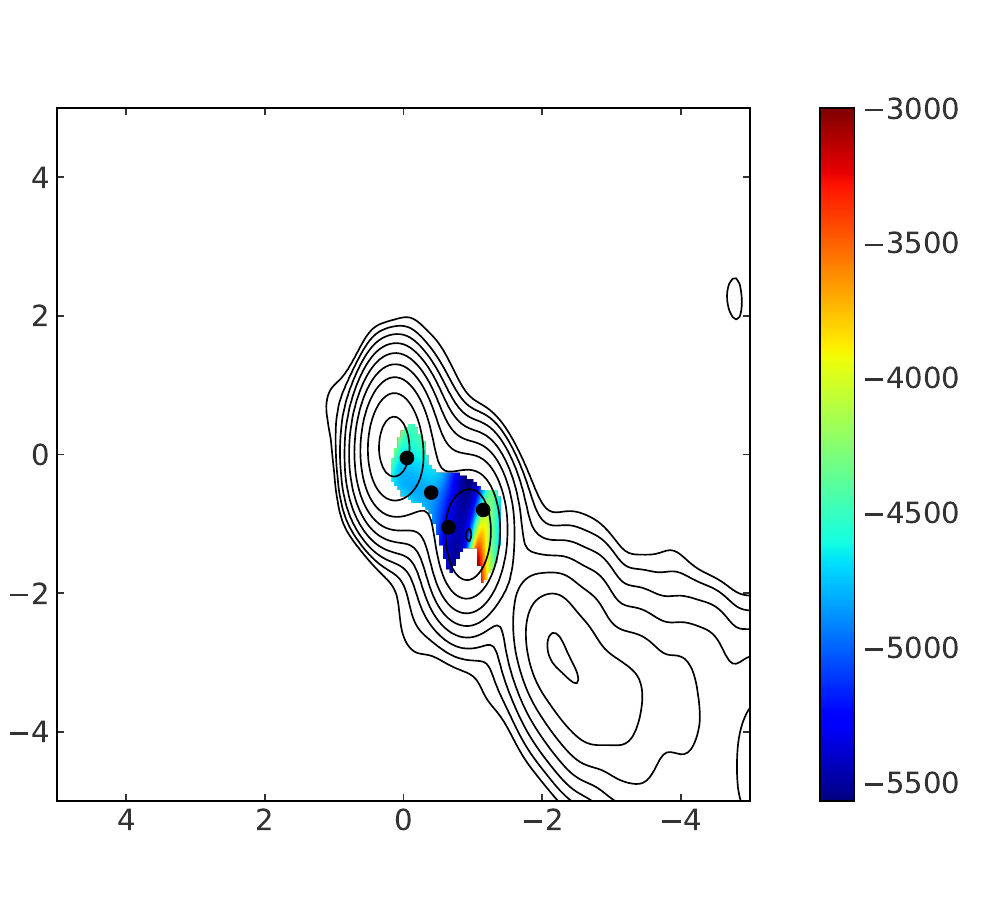}

    \caption{Linear fit of 4 pixels in the RM maps, from left to right.}
    \label{fig:pixels-fit}
\end{figure*}

\begin{figure*}[ht!]
    \centering
    \includegraphics[width=0.49\linewidth]{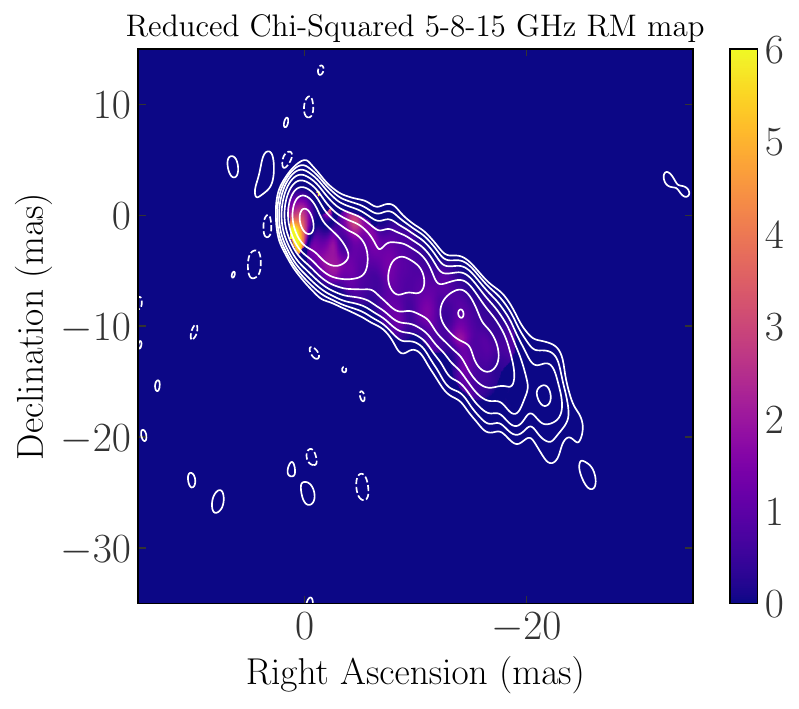}
    \includegraphics[width=0.49\linewidth]{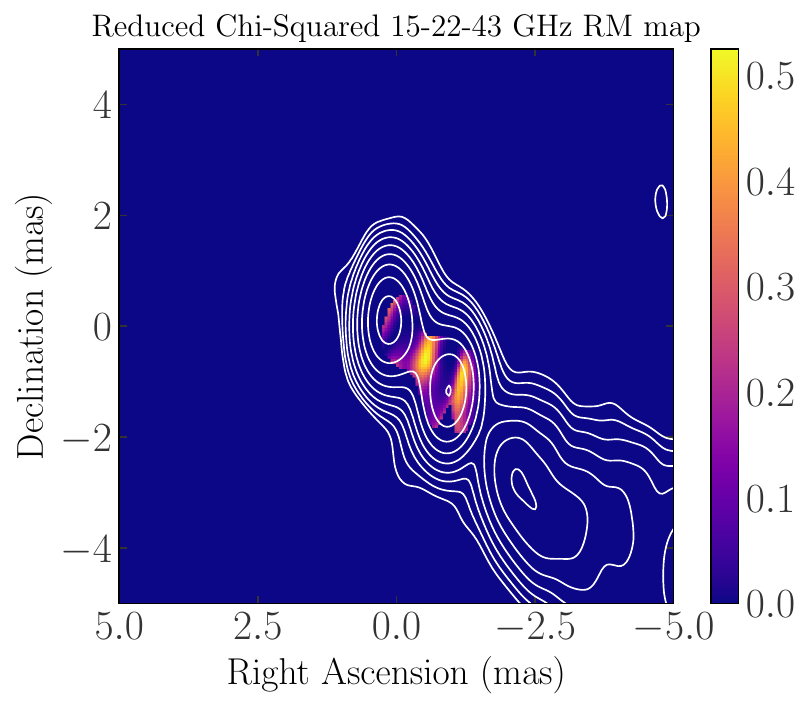}

    \caption{Goodness of fit using reduced chi squared values of both RM maps.}
    \label{fig:RM_fit}
\end{figure*}

\end{document}